# A Conceptual Framework for ERP Evaluation in Universities of Pakistan

**Sehrish Nizamani**
Department of Information Technology
Sindh University Campus @ Mirpurkhas
Email: sehrishbasir@gmail.com
Phone: +92-334-2621441

**Khalil Khoumbati**
College of Public Health and Health Informatics
Qassim University, Saudi Arabia
Email: khalil.khoumbati@gmail.com
Phone: +96-633-50414

**Imdad Ali Ismaili,**
Sindh University Campus @ Mirpurkhas
Email: iai_a@yahoo.com
Phone: +92-333-2108983

**Saad Nizamani**
Department of Information Technology
Sindh University Campus @ Mirpurkhas
Email: saad.niz@gmail.com
Phone: +92-333-2600826

**Abstract**

The higher education has been greatly impacted by worldwide trends. In a result, the universities throughout the world are focusing to enhance performance and efficiency in their workings. Therefore, the higher education has moved their systems to Enterprise Resource Planning (ERP) systems to cope with the needs of changing environment. However, the literature review indicates that there is void on the evaluation of success or failure of ERP systems in higher education Institutes in Pakistan. In overall, ERP systems implementation in higher education of Pakistan has not been given appropriate research focus. Thus, in this paper the authors have attempted to develop a conceptual framework for ERP evaluation in universities of Pakistan. This seeks to expand the knowledge on ERP in higher educational institutes of Pakistan and focuses on understanding the ERP related critical success factors.

# 1. Introduction

Enterprise Resource Planning Systems are software packages that provide the complete integration of information of various functional processes (departments) within an organization. These systems integrate with all organizational resources and shared benefits are distributed to all departments (Garcia-Sanchez and Pe´rez-Bernal, 2007). Generally, the total resources in an organization are integrated through ERP. Oracle and SAP (Systems Applications and Products) are major producers of ERP.

The Universities has turned their systems to ERP systems to meet the changing environment's needs (McCredie and Updegrove, 1999). In result of this, the legacy and other information systems are replaced or integrated with ERP system in various universities throughout the world to achieve greater efficiency and improved end-user efficiency (Kvavik et al, 2002). In addition, ERP systems are being recognized a solution to integrate academic and administrative services of universities (Rico, 2004).

In Pakistan ERP implementation process is at very low level. Higher Education Commission (HEC) of Pakistan has signed an agreement with Oracle PeopleSoft Campus Solution provider to help the public-sector universities in Pakistan with high-tech Information Technology (IT) solutions. The annual report of year 2008 by Higher Education Commission shows that the implementation of campus management solution is initially launched or implemented at seven universities of Pakistan as a pilot project. However, there is void in literature on evaluation for the implementation of ERP systems in Pakistan. Thus, in this paper the authors have made an attempt to propose the conceptual framework for the evaluation of ERP implementation.

This paper begins with the overview of ERP and emphasizes the ERPs in the Higher Education Sector. Following this, discussion is provided on ERP implantation in Universities of Pakistan, where ERP is currently active and working. Subsequently, the aim of this research is discussed. Finally, an in-depth literature review on the IS evaluation frameworks is presented and finally the conceptual framework for ERP is proposed.

## 2. Enterprise Resource Planning Systems Overview

Enterprise resource planning systems went through a range of development cycles since its beginning in 1970s as shown in Figure 1. In 1960, manufacturing systems were introduced in companies for the inventory management. In 1970s, the focus in most organizations was shifted to material requirements planning (MRP) systems. Material Requirement Planning were the most basic computerized information systems.

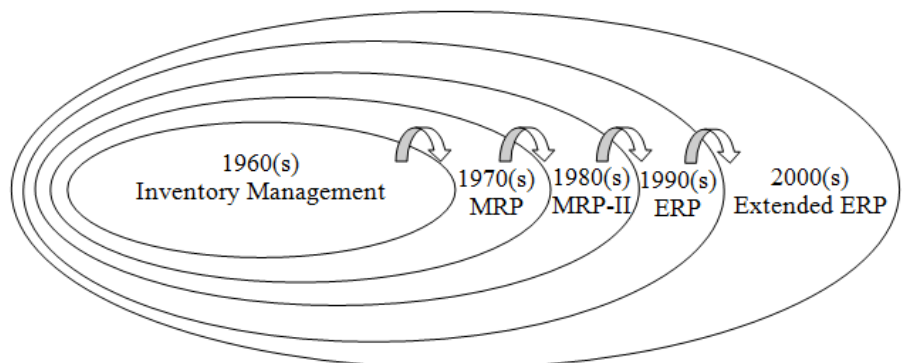

Figure 1: The Evolution of ERP

In 1980s, MRP systems extended and shifted to its second phase whereby Manufacturing Resources Planning (MRP II) developed for organizing manufacturing processes and distribution management activities (Abdinnour-Helm et al, 2003). This was a completely different approach as compared with the previous MRPs (Kakouris and Polychronopoulos, 2005).

As early as the 1990s, MRP-II started overlaying in business, engineering, finance, human resources and project management, and covered other larger areas in business; this extension of MRP-II is now known as the ERP.

Today, a new generation of ERP systems is introduced. These are termed as extended ERP systems. These systems are more advanced and more efficient in processing of order processing, procurement, sales, human resources, manufacturing, finance, accounting, sales, operations planning, customer relationship management, materials management and inventory management.

In literature, many authors have defined enterprise resource planning in different ways, Based on literature (Holland and Light, 1999; Esteves and Pastor, 2001; Davenport, 1998; Markus et al, 2000; Kumar et al, 2000; Shanks, 2000; Shang and Seddon, 2002; and Nah et al, 2001)

ERP can be defined as extensive software that comprised of multiple configurable modules integrated in a single system. As a result of that ERP system connects an organization's strategy, business processes and structure with information technology.

There are some key features of ERP such as:

a. **A common data set:** A single data set used in all of the company's internal business processes (Davenport, 1998).
b. **Standardized data definitions:** ERP business processes defined in the ERP application modules sharing the same data definitions.
c. **System adaptability:** ERP system is adaptable to enterprise changing needs (Yen et al, 2002).
d. **Outside the organization's range:** ERP systems support the online communication with the environment outside the boundary of the company and should not only be limited to the boundaries of the company.

## 3. ERP Implementation in Higher Educational Institutions

Higher education worldwide is powerfully influenced by IT development globally, especially in universities around the world due to the government's call to improve its performance and efficiency (Allen and Kern, 2001). The competitive educational environment and the expectations from the students around the world are forcing universities to improve their overall performance (Fisher, 2006). In this regard, the higher education institutions have turned themselves into the Enterprise Resource Planning systems to help them handle with the changing university environment. Accordingly, standalone applications designed for academic and administration departments were replaced by ERP in these universities (Pollock and Cornford, 2004).

University is different from other categories of business because they have different circumstances and conditions; and the ERP systems are there to fulfill the academic needs (Mehlinger, 2006). The teachers and students need vast information and improved e-learning environment. The purpose of implementing ERP in universities is to provide institution with a greater capacity for research and education (Watson and Schneider, 1999).

For universities, adaption of ERP systems is often one of the biggest challenges and a significant budget is assigned to it. Though, modest research has been carried out related to implementation of ERP in university environment judged against to other environments (Nielsen, 2002).

The main advantages of ERP for higher education include (King 2002):

- access to information for planning and improving the management of the institution
- to improve services to faculty, other staff and students
- reduce business risks
- better management of university data
- accurate and efficient data retrieval

Though it offers many advantages but ERP itself does not provide a competitive advantage and can only be used as a major supportive tool. The main priority must be the quality of service provided to staff and students in the university.

The ERP systems implementation in higher education has been considered as a difficult business. Universities often use more than $20 million each to implement ERP, and it may take at least two or more years to complete. In actual these systems were formerly designed for commercial organizations, and minor efforts have been taken to fit them to universities requirement (Beekhuyzen et al, 2001). As ERP is designed into certain modules, this is problematic for universities to adopt these packaged systems because somewhat institutions need to alter their business processes to fit into these systems (Von Hellens et al, 2005).

Enterprise Resource Planning implementation of higher education is being considered very complex project in universities environment. Therefore, the university administrative staff and other stakeholders must know about the ERP implementation issues.

## 4. ERP in Universities of Pakistan

The Higher Education Commission of Pakistan has been trying to improve performance of universities for the last several years. HEC has signed an agreement with Oracle PeopleSoft Campus Solution provider to help the public-sector universities in Pakistan with high-tech Information Technology (IT) solutions. The annual report of year 2008 by Higher Education Commission shows that the implementation of campus management solution is initially launched or implemented at six (6) universities of Pakistan as a pilot project. The names of universities included in this pilot project are as under: (Oracle PeopleSoft CMS, 2009):

- Dow University of Health Sciences, Karachi (DUHS)
- University of Engineering and Technology, Peshawar (UET)
- Quaid-e-Azam University, Islamabad (QAU)
- Islamia University, Bahawalpur (IUB)
- Balochistan University of Information Technology and Management Sciences, Quetta (BUITEMS)
- University of Punjab, Lahore (PU)

This project was started in the mid of 2007. Later on, in year 2009, the Institute of Business Administration (IBA), Sukkur was added as 7$^{th}$ Institute (HEC, 2009). In Pakistan, this project is supported by Oracle Partner Company Technologix (Oracle PeopleSoft, 2009).Three (3) other universities invested their own budget/funds and implemented ERP systems with the help of Oracle PeopleSoft Campus Solutions, namely:

- Lahore University of Management Sciences, Lahore (LUMS)
- Liaqat University of Medical and Health Sciences, Jamshoro (LUMHS)
- The Aga Khan University, Karachi (AKU)

For the universities of Pakistan, it is considered as an important shift from standalone legacy systems to campus-wide integrated solution.

Oracle is an international company with a wide range of products. 48% of US public/private-sector universities have implemented PeopleSoft solutions (Oracle PeopleSoft CMS, 2009). The solution offers variety of modules under the brand PeopleSoft Enterprise Campus Solutions.

## 5. Problem Statement

In Pakistan, the adoption of ERP system is in its early stage. Major international vendors of ERP solutions are strongly marketing and offering their products to developing countries (Pairat and Jungthirapanich, 2005). In many developing countries such as Pakistan, the implementation of ERP systems is being considered a very complicated issue. In many studies, along with (Huang and Palvia, 2000; Upadhyay et al, 2010), it has been found that limited capital amount, lack of experience in IT and non-availability of resources influence the ERP implementation in developing countries.

It is also reported in literature (Richtermeyer and Bradford, 2002; El Sawah et al, 2008) that there is void on evaluating the ERP implementations, particularly in developing countries. Rabaa'i et al (2009) also raises questions regarding the effective evaluation of ERP in higher education sector. It is, therefore, very significant to determine the success of ERP implementations, because a huge budget and human resources are invested therein.

However, there is limited empirical research stated on the evaluation of ERP in public or private universities of Pakistan. Although, some authors such as (Anjum et al, 2010; Shah et al, 2011) have recognized some critical success factors for ERP implementation in industries of Pakistan, yet, they did not clarify the success or failure of the implementation of ERP in those industries. Thus, this study attempt to propose a model for evaluation of ERP systems in higher educational institutes of Pakistan.

## 6. A Conceptual Framework for ERP Evaluation

Miles and Huberman (1994) suggest that a theoretical framework elaborates the important issues to be studied. Several authors proposed their research agenda in evaluating the Information Systems and ERP success models (DeLone and McLean, 1992; Pitt et al, 1995; Seddon, 1997; DeLone and McLean, 2003).

The literature provides diverging and fluctuating results regarding which variables are required for the evaluation of successful ERP implementation. In brief, there is no single factor or set of factors for evaluation of successful ERP implementation. The combination of factors helps analyze the success. These factors change over time over circumstances or over the nature of business.

Thus, for the purpose of this research authors have reviewed various models for ERP implementation success and analyzed several factors for successful ERP implementation. The two broad categories of research areas from which variables are analyzed for evaluating ERP implementation includes:

- The success dimensions (treated as dependant variables) from the model of evaluation of IS implementation (DeLone and McLean, 1992).

- Critical success factors (treated as independent variables) that ensure the successful ERP implementation in higher education institutions of Pakistan.

The DeLone and McLean (1992) model is the widely cited and most known model for evaluating information systems success (Heo and Han, 2003; Urbach et al, 2009) and it is recently determined as the most cited article from year 1992 till year 2007 in the top journals of information systems (Petter et al, 2013). This model is considered as the standard for justifying the dependent variables for IS success (Myers et al, 1997). In their research article, the authors studied 180 papers in theoretical and experimental studies of success factors for Information Systems in which they categorized evaluation process in six broad categories i.e. system quality, information quality, use, user satisfaction, individual impact and organizational impact as shown in figure 2.

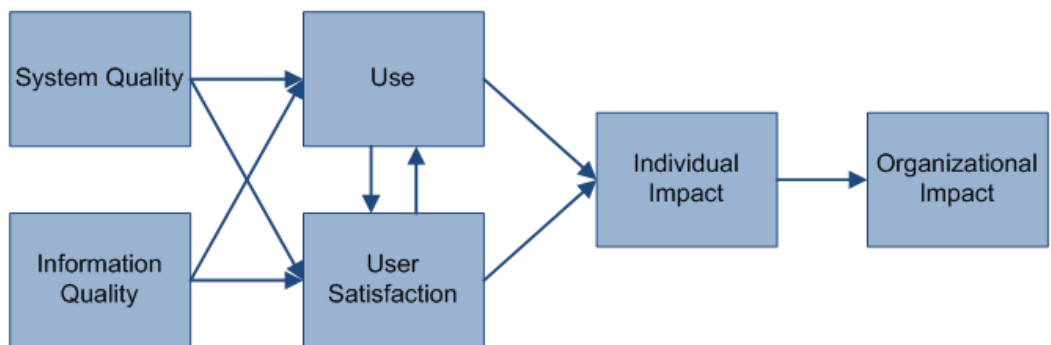

**Figure 2: The Model of DeLone& McLean, 1992**
Source (DeLone and McLean, 1992)

Based on the literature of information system success model proposed by DeLone and McLean (2003) and the critical success factors that affect the successful ERP implementation, a conceptual framework is proposed in this research. From DeLone and McLean (2003) model, the dependant variables are defined. As DeLone and McLean models are widely accepted in literature, so it will help us to measure ERP success through different angles.

In order to meet the characteristics of ERP systems for the higher education sector of Pakistan, the measures or success dimensions have been revised from the model of information system success proposed by DeLone and McLean (1992) as follows:

a. Elimination of 'Use' construct: The ERP systems are mandatory to use by the end users in universities. There is a little variability of Use construct in systems which are mandatory to use, hence this construct can be eliminated (Petter et al, 2008; Sedera and

Gable, 2004; Seddon, 1997). After the justification and clarification of this construct by DeLone and McLean (2003), the construct Use is still questionable in recent literature (Bradley et al, 2006; Gable et al, 2003).

b. Inclusion of 'Service Quality' construct: There is more dependency on vendors and consultants in developing countries as compared with developed countries (Asemi and Jazi, 2010) because these systems are designed in the countries which are entirely different in culture with those countries where these systems are implemented. The service quality variable for the support in e-commerce based systems was also added by DeLone and McLean (2003) by admitting the criticism of Pitt et al (1995).

The revised model is purely based on review of literature and is represented by six dimensions. In terms of measuring ERP implementation success, these dimensions are selected as indicators of the dependent variable. In the figure 3, the System Quality, Information Quality and Service Quality singularly and mutually affect User Satisfaction. User Satisfaction directly affects Individual Impact. Individual Impact should eventually affect Organizational Impact.

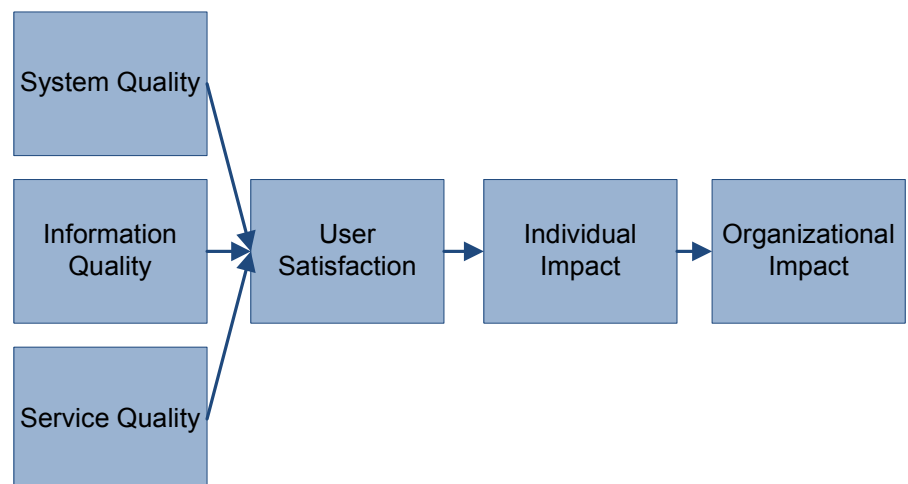

**Figure 3: Proposed ERP Success Dimensions**

The literature is reviewed regarding what critical success factors are important for ERP in developing countries and in higher education as well. For this research, only four critical success factors are selected. These include Top Management Support, Business Process Reengineering, Organization Culture, Education and Training. The selected CSF served as independent variables. These factors are selected as they are highly reported in literature as compared with other factors with ERP implementations in higher education, in developing countries and CSF in general. The research framework is depicted in figure 4.

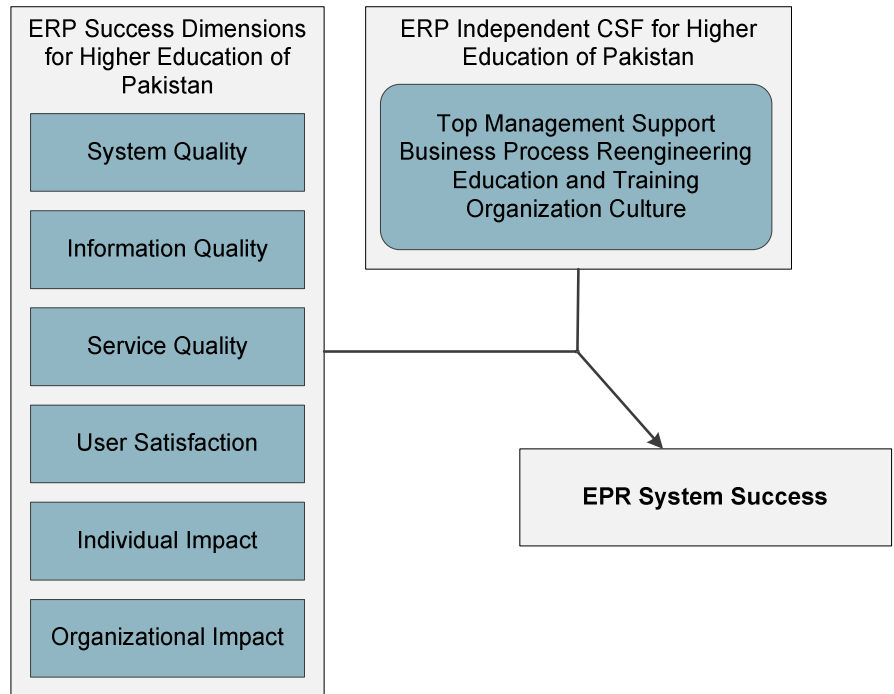

Figure 4: Proposed ERP Success Model

ERP Success Dimensions include System Quality, Service Quality, Information Quality, User Satisfaction, Individual Impact and Organizational Impact. The CSFs include Top Management Support, Business Process Reengineering, Organization Culture and Education and Training. The dependent variables (ERP success dimensions) and independent variables (CSFs) have jointly affected the success/failure of ERP in Higher Education of Pakistan. These variables may have positive or negative impact.

### 6.1. System Quality

System Quality is a measure of how appropriate the software and hardware work together. It is a multi-faceted construction variable proposed to understand the work of system by technically and by means of design (Gable et al, 2008). System Quality is considered as one of the extensively studied factor according to DeLone and McLean (1992).

### 6.2. Information Quality

According to Pitt et al (1995), the Information Quality refers to a measure of determining the quality of output of information system and has its own importance for the ERP end users (Ng, 2001). The modules of the ERP system are closely related to each other, inaccurate data entry in a module affects the operation of other modules. This can also be explained as garbage in garbage out. Therefore, the information quality is an important factor for ERP implementation successfully (Yusuf et al, 2004).

**6.3. Service Quality**

Several authors argue that the information system success models must include the Service Quality construct as most of the budget devoted to ERP services. As a result, a new construct named "Service Quality" has been introduced by adapting the measuring instrument SERVQUAL from business context (DeLone and McLean, 2003; Kettinger and Lee, 2005). This dimension captures the quality of services that a particular information system provides to an organization (DeLone and McLean, 2003).

**6.4. User Satisfaction**

According to DeLone and McLean (1992), the dependent variable widely used to measure the performance of the system found in literature is user satisfaction. It is the received response generated through the use of information system. If players are more willing to participate in the information and data sharing activities throughout the system, it results in higher level of user satisfaction with ERP systems. It is also considered as the necessary condition for repeated use of data.

**6.5. Individual Impact**

The individual impact can be defined as the effect of IS to the behavior of end-user (DeLone and McLean, 1992). Individual impact in itself is the most ambiguous to define. The authors describe how each effect is a general term used to reflect how the information affects the user. In this study, the individual impact is restated as the effect of the ERP system on the activities and performance of a user of ERP.

**6.6. Organizational Impact**

According to DeLone and McLean (1992), organizational impact is defined as a construct designed to concentrates on the effectiveness of ERP in affecting organizational performance (DeLone and McLean, 1992). These measures can be grouped into three different areas: studies with the profits, productivity studies and studies using cost / benefit analysis. From these three areas, one may select one or more measures to analyze organizational impact.

### 6.7. Top Management Support

Top management of an organization must support ethically, morally, financially and through the provision of resources to achieve the goals in time (Sabau et al, 2009). For the successful implementation of ERP systems, the commitment of senior management is accepted as one of the key factors (Bhatti, 2005). In addition, it is the most quoted CSF for ERP implementation. The major responsibility for senior management is to assist financially in order to build the successful system. Top management must ensure the people that this project is highly prioritized in organization. The lack of enough resources and financial support leads to ERP implementation failure or will lack in providing full range of benefits (Beheshti, 2006). Senior management should possess strong leadership in order to demonstrate its commitment towards the ERP project, analyze the scope of project and achieve success in its implementation

### 6.8. Business Process Re-engineering

Shehab et al (2004) explains that in order to implement ERP system, one of the two strategic approaches is to be selected by an organization. In the first method, organizations need to redesign business processes in order to adapt to the functionality of modules of ERP. This involves changes to core business processes; organizations carry out the business tasks in improved manner, and the responsibility of the employees gets modified (such as: Hong and Kim, 2002; Holland and Light, 1999). Another approach is to customize the ERP software package in order to fit that software in business processes. Customization in the modules should be avoided or minimized to achieve maximum functionality of ERP system (Shanks et al, 2000).

### 6.9. Education and Training

This factor summarizes user training, extensive training, training in business processes, user participation in user education and proper training. Training must show employees why ERP is important and implemented. Employees must understand the relation of their work to other functional areas of business (Zhang et al, 2003). Appropriate training programs should be planned and offered to end users of the new system (Bajwa et al, 2004). Furthermore, the provision of user manuals in print and online tutorials, workshops and assistance can be utilized to ensure proper understanding of system and to support users. The appropriate user training is accepted as a major factor in successful ERP systems implementation (Mabert et al, 2003; Al-Mashari et al, 2003; Somers and Nelson, 2001).

## 6.10. Organization Culture

Organization Culture is the most important factor in developing countries (Asemi and Jazi, 2010). This factor plays a significant role (Zhang et al, 2005). The organization culture can be defined as "a pattern of shared basic assumptions that the group learned as it solved its problems of external adaptation and internal integration. It has worked well enough to be considered valid and, therefore, to be taught to new members as the correct way to perceive, think, and feel in relation to those problems" (Schein, 2006). Organizational culture refers various aspects that include the processes of the organization, personal values of employees, decision making processes, attitudes and skills of employees.

## 7. Research Model and Summary of Hypothesis

The table 1 provides a complete summary of hypothesis formulated from the conceptual research model; whereas; Figure 5 represents the source of formulation of those hypotheses by directional arrows.

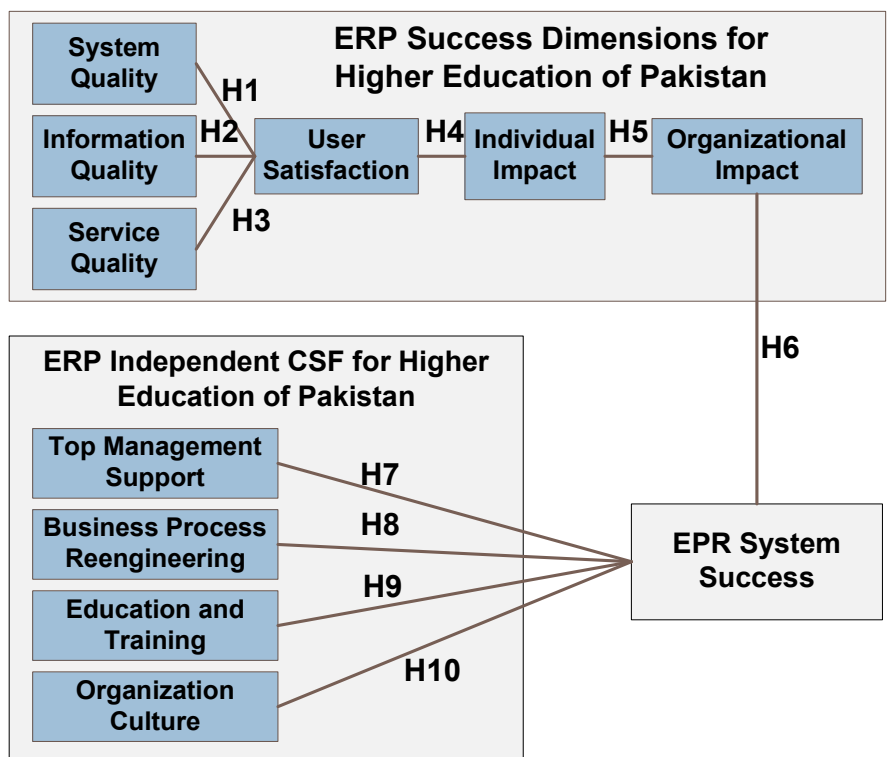

**Figure 5: Research Model and Hypothesis**

| Success Dimensions / CSF | Hypothesis |
|---|---|
| System Quality, Information Quality and Service Quality singularly and mutually affect User Satisfaction. | H1: System Quality is positively or negatively related to user satisfaction. |
| | H2: Information Quality is positively or negatively related to user satisfaction. |
| | H3: Service Quality is positively or negatively related to user satisfaction. |
| User Satisfaction directly affects Individual Impact. | H4: User Satisfaction is positively or negatively related to Individual Impact. |
| Individual Impact eventually affects Organizational Impact. | H5: Individual Impact is positively or negatively related to organizational impact. |
| Organizational Impact leads to ERP System Success or Higher Education of Pakistan. | H6: Organizational Impact is positively or negatively related to ERP System Success. |
| Top Management Support, Business Process Reengineering, Education and Training and Organization Culture singularly and mutually affect success of ERP Systems in Higher Education of Pakistan. | H7: Top Management Support is positively related to ERP System Success. |
| | H8: Business Process Reengineering is positively related to ERP System Success. |
| | H9: Education and Training is positively related to ERP System Success. |
| | H10: Organization Culture is positively related to ERP System Success. |

Table 1: Summary of Hypothesis

## 8. Materials and Methods

The survey research method is one of the most leading research methods used in information system research (Pinsonneault and Kraemer, 1993) in order to determine dependant and independent variables of an environment without having any control on it.

The dependent and independent variables elaborate the area of study in survey research. These variables are not controlled by researcher explicitly. Before performing survey, a model is designed that includes dependant or independent variables and relationship between those variables. The survey questionnaire is constructed based on those variables and their interrelationships. The observations are gathered to test the model. Survey presents a snapshot at particular time. Moreover, it acts as a verification tool (Gable, 1994).

This survey is planned and designed as a cross-sectional survey. Information is collected from respondents while considering a single time and not longitudinal in nature.

The online surveys are used as the medium for this research. The survey questionnaires are distributed to following targeted universities through electronic mails.

- Dow University of Health Sciences, Karachi
- University of Engineering and Technology, Peshawar
- Quaid-e-Azam University, Islamabad
- Islamia University, Bahawalpur
- Balochistan University of Information Technology and Management Sciences, Quetta
- University of Punjab, Lahore
- Institute of Business Administration, Sukkur
- Lahore University of Management Sciences (LUMS)
- Liaqat University of Medical and Health Sciences (LUMHS)
- The Agakhan University (AKU)

The above mentioned universities are identified because in these universities ERP systems (named as Campus Management Solutions) is implemented, active and working. The target population includes all those from the above population who access the CMS to perform their university tasks. The exclusion criteria exclude students and all others.

## 9. Conclusions

In this paper a conceptual framework is proposed to establish the most important factors to research for implementing an ERP system in higher education institutions of Pakistan. This conceptual framework is developed after reviewing of a literature on the existing quality frameworks and existing models. The critical success factor for ERP implementations are also taken into account.

The theoretical model is developed comprising six dimensions named System Quality, Information Quality, Service Quality, User Satisfaction, Individual Impact and Organizational Impact and the four critical success factors namely Top Management Support, Business Process Reengineering and Organizational Culture; Education and Training. In the future research this model will be empirically tested to validate these factors.

Z. ZHANG, M.K. LEE, P. HUANG, L. ZHANG, and X. HUANG (2005) A framework of ERP systems implementation success in China: an empirical study. *International Journal of Production Economics,* **98**(1), pp. 56-80.